\documentclass[12pt]{article}

\begin{document}

\title{Ginsburg-Landau Expansion in a non-Fermi Superconductor}
\author{C. P. Moca\footnote{Corresponding author. Fax +40-59-432789;E-mail:mocap@ff.uoradea.ro} \\
Department of Physics, University of Oradea\\
3700, Oradea, Romania}
\maketitle

\begin{abstract}
We study the Ginsburg-Landau expansion for the non-Fermi model proposed by
Anderson. We analyze the deviations of the main properties of a non-Fermi
superconductor from the isotropic s-wave bidimensional superconductor. \\%
\end{abstract}

\section{Introduction}

Since the discovery of high temperature superconductivity (HT$_c$S) the
microscopic description of the HT$_c$S was not a simple problem. It is
generally accepted that the normal and superconducting phase are not
describe by a Fermi theory. There are a lot of microscopical models
describing the normal and superconducting phase. The nearly
antiferromagnetic model proposed by Millis, Monien and Pines \cite{1},
non-Fermi Anderson model \cite{2}, marginal Fermi liquid proposed by C.M.
Varma et.al. \cite{3} are phenomenological models used to explain the normal
state physical properties of high temperature superconductors. Using the
BCS-like model, the Gorkov equations have been applied to describe the
superconducting state in a non-Fermi liquid, proposed by the Anderson model 
\cite{2}. The superconducting state properties have been studied by
different authors \cite{4,5}. The idea of this model is the existence of a
state in bidimensional systems similar with the state described by the
Luttinger liquid for one-dimensional systems. In such a system the spectral
function $A(\mathbf{k},\omega )=-ImG(\mathbf{k},\omega )$ satisfies the
homogeneity relation $A(\Lambda \mathbf{k},\Lambda \omega )=\Lambda
^{-1+\alpha }A(\mathbf{k},\omega )$ with an exponent $\alpha $ greater than
zero [5]. This spectral anomaly implies the break down of the Fermi liquid
theory, the limit $\alpha =0$ is a special choice leading to the Fermi
liquid model. Wen [12] has demonstrated that the exponent $\alpha $ is
non-universal depending on coupling between the fermions.

In this paper we analyze the properties of the non-Fermi superconductor
using the method developed in Ref.\cite{6}. We calculate the Ginsburg-Landau
coefficients ( Sec. III ) and analyze the main physical properties of the
non-Fermi superconductor (Sec. IV). Finally we compare the results with
other theoretical models ( Sec. V ) for the cuprates superconductors.

\section{Model}

The non-Fermi behavior of the normal state of cuprates superconductors has
been discussed by different authors \cite{4,5} in order to explain the
physical properties of the cuprate superconductors. In the following we
consider the superconducting phase which is described by the BCS-like order
parameter $\Delta _{\mathbf{k}}$ which can be calculated from the Gorkov
equations. In the superconducting phase the electrons are described by the
normal and anomalous Green functions: 
\begin{equation}
G(\mathbf{k},\omega )=e^{-i\pi (1-\frac \alpha 2)sgn\left( \omega \right) }%
\frac{\left( \frac{g(\alpha )}{\omega _c^\alpha }\right) ^{-1}\left( \omega
+\varepsilon _{\mathbf{k}}-i\delta \right) ^{1-\alpha }}{\left( \frac{%
g(\alpha )}{\omega _c^\alpha }\right) ^{-2}\left( \varepsilon _{\mathbf{k}%
}^2-\omega ^2-i\delta \right) ^{1-\alpha }+\left| \Delta _{\mathbf{k}%
}\right| ^2},  \label{1}
\end{equation}
\begin{equation}
F(\mathbf{k},\omega )=\frac{-\Delta _{\mathbf{k}}}{\left( \frac{g(\alpha )}{%
\omega _c^\alpha }\right) ^{-2}\left( \varepsilon _{\mathbf{k}}^2-\omega
^2-i\delta \right) ^{1-\alpha }+\left| \Delta _{\mathbf{k}}\right| ^2},
\label{2}
\end{equation}
where $\omega _c$ is a cut-off energy, $0<\alpha <1$, g($\alpha $)=$\pi
\alpha /2\sin (\pi \alpha /2)$. The equation for the transition temperature T%
$_c$ can be obtained from the equation for the Cooper instability: 
\begin{equation}
1-\Pi (\mathbf{0},0)=0,  \label{3}
\end{equation}
where the Cooper susceptibility $\Pi (0,0)$ can be calculated following Ref.%
\cite{6}. As a result the equation for T$_c$ is : 
\begin{equation}
\frac 1V=T_c\sum\limits_n\int \frac{d^2\mathbf{k}}{\left( 2\pi \right) ^2}%
\left[ G(\mathbf{k},i\omega _n)G(-\mathbf{k},-i\omega _n)+F(\mathbf{k}%
,i\omega _n)F(-\mathbf{k},-i\omega _n)\right] ,  \label{4}
\end{equation}
where V is the attractive interaction between the electrons in the
superconducting state. The critical temperature T$_c$ was also calculated in
Ref.\cite{7} and the final result is: 
\begin{equation}
T_c=\omega _D\left( \frac{D\left( \alpha \right) }{C\left( \alpha \right) }%
\right) \exp \left[ -\frac 1V\frac 1{2\alpha A\left( \alpha \right) D\left(
\alpha \right) }\left( \frac{\omega _c}{\omega _D}\right) ^\alpha \right] ,
\label{5}
\end{equation}
where $A(\alpha )$, $C(\alpha )$, $D(\alpha )$ are functions of $\alpha $
given by Ref.\cite{7}. The critical temperature T$_c$ was calculated also in
Ref.\cite{8}. The small difference from the result obtained in Ref.\cite{7}
has the origin in the mathematical approximations made in the calculations.
In deriving eq. (5) the sum over Matsubara frequencies is calculated first
and than the integration over the momentum.

The result (\ref{5}) gives in the limit $\alpha \rightarrow 0$ the well
known result for T$_c$: 
\begin{equation}
T_c=1.13\omega _D\exp \left[ -\frac 1{N(0)V}\right] ,  \label{6}
\end{equation}
which make us to believe that eq.(\ref{5}) may be correct. The result given
in Ref.\cite{8} also in the limit $\alpha \rightarrow 0$ gives the BCS
result for the critical temperature.

\section{Ginsburg-Landau Expansion}

The difference between the superconducting and normal free energy densities
is written in the usual Ginsburg-Landau expansion: 
\begin{equation}
F_S-F_N=a\left| \Delta _{\mathbf{k}}\right| ^2+\frac b2\left| \Delta _{%
\mathbf{k}}\right| ^4+c\mathbf{k}^2\left| \Delta _{\mathbf{k}}\right| ^2,
\label{7}
\end{equation}
where $a$, $b$, $c$, are the Ginsburg-Landau coefficients. The calculation
of the critical coefficients is done following Ref.\cite{6,13}. The
coefficient $a$ is given by the relation: 
\begin{equation}
a=\frac 1V-T\sum\limits_n\int \frac{d^2\mathbf{k}}{\left( 2\pi \right) ^2}%
\left[ G(\mathbf{k},i\omega _n)G(-\mathbf{k},-i\omega _n)+F(\mathbf{k}%
,i\omega _n)F(-\mathbf{k},-i\omega _n)\right] .  \label{8}
\end{equation}
Evaluation in eq. (\ref{8}) was done using eq. (\ref{4}) for the attractive
potential between the fermions. The integration over $\mathbf{k}$ is
transformed using the relation $\int \frac{d^2\mathbf{k}}{\left( 2\pi
\right) ^2}\rightarrow N(0)\int d\varepsilon $ where $N(0)$ is the density
of states. After a short calculation we get for $a$ the following
expression: 
\begin{equation}
a=2\pi ^{2\alpha -1}N\left( 0\right) \left( \frac{g\left( \alpha \right) }{%
\omega _c}\right) ^2\left( T^{2\alpha }-T_c^{2\alpha }\right)
\int\limits_0^\infty \frac{dx}{\left( 1+x^2\right) ^{1-\alpha }}%
\sum\limits_n^N\left( 2n+1\right) ^{2\alpha -1},  \label{9}
\end{equation}
where $x=\varepsilon /\omega _n$ is a new parameter and $\omega _n=(2n+1)\pi
T$ are the fermionic Matsubara frequencies. $N$ is connected with the cutoff 
$\omega _D$ by the relation $N=\omega _D/\pi T_c$. After performing the
integration over $x$ and the summation over $n$ we get the final result: 
\begin{equation}
a=N\left( 0\right) \frac{T-T_c}{T_c}\left( \frac{\omega _D}{\omega _c}%
\right) ^{2\alpha }g^2\left( \alpha \right) \frac{B\left( \frac 12,\frac
12-\alpha \right) }\pi ,  \label{10}
\end{equation}
where T$_c$ is the critical temperature given by eq.(\ref{5}) and $%
B(x,y)=\Gamma \left( x\right) \Gamma (y)/\Gamma (x+y)$, and $\Gamma \left(
x\right) $ are Euler functions. In the limit $\alpha \rightarrow 0$ we
obtain for the coefficient $a$: 
\begin{equation}
a_0=N\left( 0\right) \frac{T-T_{co}}{T_{co}},  \label{11}
\end{equation}
which is the expression for the case of a two-dimensional isotropic
superconductor. The coefficient $b$ can be written as \cite{6,13}: 
\begin{equation}
b=T_c\sum\limits_n\int \frac{d^2\mathbf{k}}{\left( 2\pi \right) ^2}\left[ G(%
\mathbf{k},i\omega _n)G(-\mathbf{k},-i\omega _n)+F(\mathbf{k},i\omega _n)F(-%
\mathbf{k},-i\omega _n)\right] ^2.  \label{12}
\end{equation}
The calculation for evaluating $b$ follows the same procedure described for
the calculation of $a$, the only differences being the appearence of
different exponents in the $x$ integration and summation over frequencies.
After performing the calculations in eq.(\ref{12}) we obtain for the $b$ the
result: 
\begin{equation}
b=N\left( 0\right) \frac{7\zeta (3)}{8\pi ^2T_c^2}g^4\left( \alpha \right)
\left( \frac{2\pi T_c}{\omega _c}\right) ^{4\alpha }\frac{2B\left( \frac
12,\frac 32-2\alpha \right) }\pi \frac{2^{3-4\alpha }-1}7\frac{\zeta \left(
3-4\alpha \right) }{\zeta \left( 3\right) },  \label{13}
\end{equation}
where $\zeta \left( x\right) $ is Riemann function. In the limit $\alpha
\rightarrow 0$ we obtain the well known result: 
\begin{equation}
b_0=N\left( 0\right) \frac{7\zeta (3)}{8\pi ^2T_{co}^2}.  \label{14}
\end{equation}
In order to calculate $c$ we must perform a Taylor expansion in powers of $%
\mathbf{q}$ of the expression: 
\begin{equation}
-T_c\sum\limits_n\int \frac{d^2\mathbf{k}}{\left( 2\pi \right) ^2}\left[ G(%
\mathbf{k+}\frac{\mathbf{q}}2,i\omega _n)G(-\mathbf{k+}\frac{\mathbf{q}}%
2,-i\omega _n)+F(\mathbf{k+}\frac{\mathbf{q}}2,i\omega _n)F(-\mathbf{k+}%
\frac{\mathbf{q}}2,-i\omega _n)\right]   \label{15}
\end{equation}
The integration over $\mathbf{k}$ is now transformed using the relation $%
\int \frac{d^2\mathbf{k}}{\left( 2\pi \right) ^2}\rightarrow N(0)\int
d\varepsilon \int\limits_0^{2\pi }\frac{d\theta }{2\pi }$ because of the
appearence of the $\cos (\theta )$ term where $\theta $ is the angle between 
$\mathbf{k}$ and $\mathbf{q}$. We have to consider the first and the second
orders of Taylor expansion in order not to loose some important
contributions. After performing the calculations in eq.(\ref{15}) we get for 
$c$ the result: 
\[
c=N\left( 0\right) \frac{v_F^27\zeta (3)}{32\pi ^2T_c^2}g^2\left( \alpha
\right) \left( \frac{2\pi T_c}{\omega _c}\right) ^{2\alpha }\frac{%
2^{3-2\alpha }-1}7\frac{\zeta \left( 3-2\alpha \right) }{\zeta \left(
3\right) }
\]
\begin{equation}
\times \frac{\left( 3-\alpha \right) \left( 1-\alpha \right) B\left( \frac
12,\frac 52-\alpha \right) -\left( 1-\alpha ^2\right) B\left( \frac 32,\frac
32-\alpha \right) }\pi .  \label{16}
\end{equation}
In the limit $\alpha \rightarrow 0,$ $c$ becomes: 
\begin{equation}
c_0=N\left( 0\right) \frac{v_F^27\zeta (3)}{32\pi ^2T_{co}^2}.  \label{17}
\end{equation}
$c_0$ is the coefficient for the s-wave isotropic superconductor. From eqs.(%
\ref{11}), (\ref{14}), (\ref{17}) we can see that we reobtain the results
for the ordinary superconductors, for the Ginsburg-Landau coefficients in
the limit $\alpha =0$.

\section{Physical Characteristics of a non-Fermi Superconductor}

In this section we present the behavior of the coherence length, the
penetration depth of the magnetic field and the jump in the heat capacity at
the transition point for a non-Fermi superconductor. The coherence length at
a given temperature is given by the following expression between the
Ginsburg-Landau coefficients: 
\begin{equation}
\xi ^2\left( T\right) =-\frac ca.  \label{18}
\end{equation}
In the ordinary isotropic superconductor $\xi _0\left( T\right) =\sqrt{%
-c_0/a_0}=0.74\xi _0/\sqrt{1-T/T_{co}}\,$ where $\xi _0=0.18v_F/T_{co}$ \cite
{9}. Using eqs. (\ref{10}), (\ref{16}) we obtain for the coherence length
the expression: 
\[
\xi ^2\left( T\right) =\xi _0^2\left( T\right) \left( \frac{T_{co}}{T_c}%
\right) ^2\left( \frac{1-\frac T{T_{co}}}{1-\frac T{T_c}}\right) \left( 
\frac{2\pi T_c}{\omega _D}\right) ^{2\alpha }\frac{2^{3-2\alpha }-1}7\frac{%
\zeta \left( 3-2\alpha \right) }{\zeta \left( 3\right) }
\]
\begin{equation}
\frac{\left( 3-\alpha \right) \left( 1-\alpha \right) B\left( \frac 12,\frac
52-\alpha \right) -\left( 1-\alpha ^2\right) B\left( \frac 32,\frac
32-\alpha \right) }{B\left( \frac 12,\frac 12-\alpha \right) }.  \label{19}
\end{equation}
From eq. (\ref{19}) we find the same temperature dependence of the coherence
length $\xi \left( T\right) =\xi /\sqrt{1-T/T_c}$. At a given temperature T
the coherence length is a function of $\alpha $. The dependence of $\xi
\left( T\right) $ as a function of $\alpha $ is presented in Fig.1 .
The general expression for the penetration depth in terms of Ginsburg-Landau
coefficients is given by: 
\begin{equation}
\lambda ^2\left( T\right) =-\frac{\overline{c}^2}{32\pi e^2}\frac b{ac}.
\label{20}
\end{equation}
The general expression (\ref{20}) can be simplified if we introduce the $%
\lambda _0^2\left( T\right) $ as the penetration depth in a ordinary
superconductor \cite{9} as: 
\begin{equation}
\lambda _0\left( T\right) =\frac 1{\sqrt{2}}\frac{\lambda _0}{\sqrt{%
1-T/T_{co}}},  \label{21}
\end{equation}
where $\lambda _0^2=m\overline{c}^2/4\pi ne^2$ is the penetration depth at
T=0. Using eqs. (\ref{20}) and (\ref{21}) we have for $\lambda ^2(T)$: 
\[
\lambda ^2\left( T\right) =\lambda _0^2\left( T\right) \left( \frac{1-\frac
T{T_{co}}}{1-\frac T{T_c}}\right) \left( \frac{2\pi T_c}{\omega _D}\right)
^{2\alpha }\frac{2^{3-4\alpha }-1}{2^{3-2\alpha }-1}\frac{\zeta \left(
3-4\alpha \right) }{\zeta \left( 3-2\alpha \right) }
\]
\begin{equation}
\times \frac{2\pi B\left( \frac 12,\frac 32-2\alpha \right) }{B\left( \frac
12,\frac 12-\alpha \right) \left[ \left( 3-\alpha \right) \left( 1-\alpha
\right) B\left( \frac 12,\frac 52-\alpha \right) -\left( 1-\alpha ^2\right)
B\left( \frac 32,\frac 32-\alpha \right) \right] }.  \label{22}
\end{equation}
The temperature dependence of the penetration depth is the same as that in
ordinary superconductors, but we have an $\alpha $ dependence of $\lambda
^2\left( T\right) $ as in Fig.2.
Using eqs.(\ref{19}) and (\ref{22}) it is possible to calculate the
Ginsburg-Landau parameter $k=\lambda \left( T\right) /\xi \left( T\right) $.
The dependence of $k$ as function of $\alpha $ is the same with the $\lambda
\left( T\right) $ dependence. We can also calculate the size of the
discontinuity in the heat capacity at the transition point. The jump in the
heat capacity is given by: 
\begin{equation}
\frac{C_S-C_N}\Omega =\frac{T_c}b\left( \frac a{T-T_c}\right) ^2,  \label{23}
\end{equation}
where $C_S$ and $C_N$ represent the superconducting and normal heat
capacities, and $\Omega $ is the volume of the system. If we consider the
ratio between the jump in the heat capacity of a non-Fermi superconductor
and the jump of a normal bidimensional superconductor we obtain: 
\begin{equation}
\frac{\left( C_S-C_N\right) _{T_c}}{\left( C_S-C_N\right) _{T_{co}}}\frac{%
T_{co}}{T_c}=\left( \frac{T_c}{T_{co}}\right) ^2\left( \frac{\sqrt{\omega
_c\omega _D}}{2\pi T_c}\right) ^{4\alpha }g^{-2}\left( \alpha \right) \frac{%
B^2\left( \frac 12,\frac 12-\alpha \right) }{2\pi B\left( \frac 12,\frac
32-2\alpha \right) }\frac 7{2^{3-4\alpha }-1}\frac{\zeta \left( 3\right) }{%
\zeta \left( 3-4\alpha \right) }  \label{24}
\end{equation}
The $\alpha $ dependence is presented in Fig.3.
As we can see in Fig. 3 the jump in $C_{V}$ decreases with
increasing $\alpha $.

\section{Conclusions}

We have calculated the critical temperature and the Ginsburg-Landau
coefficients for a non-Fermi superconductor \cite{2}. We have found a
similar behavior of the Ginsburg-Landau coefficients with those in the BCS
case. The results are in agreement with previous calculations \cite{10}. A
similar model was used by Sudbo \cite{4} and Muthukumar \cite{11} for the
calculation of critical temperature. Using these results we calculated the
coherence length, the penetration depth and the jump in the specific heat at
the transition point. We have to mention that from a non-Fermi
superconductor we can obtain the results for the BCS case by putting $\alpha
=0$ in eqs.(\ref{19}), (\ref{22}), (\ref{24}). Taking some reasonable value
of the specific heat jump\cite{14,15,16} [ see Fig. 3 ] we get that $\alpha $
should satisfy the condition $0.2<\alpha <0.4$.

In high temperature superconductivity the overdoped region is well described
by the standard Fermi liquid theory. Our theory is applicable in the
underdoped and slightly overdoped regions where the Fermi liquid theory
breaks down. In these regions there is another problem which should be taken
into consideration, the opening of a pseudogap in the normal state of these
materials.

Finally we mention that these calculations give a qualitative idea on the
superconductivity in the non-Fermi superconductor. A quantitative comparison
with the experiment seems difficult because of the existence of many
parameters in the theory.

\section*{Acknowledgments}

The author is grateful to Prof. M. Crisan for his comments on the physical
aspects of the model.

\newpage\ 

\begin{center}
\textbf{Figure Captions}
\end{center}

Fig. 1 The $\alpha $ dependence of the coherence length for T=0.1T$_c$ and
T=0.8T$_c$ given by eq. (\ref{19})

Fig .2 The $\alpha $ dependence of the penetration depth for T=0.1T$_c$ and
T=0.8T$_c$ given by eq. (\ref{22})

Fig. 3 The $\alpha $ dependence of the specific heat jump at the critical
point given by eq. (\ref{24})

\ 

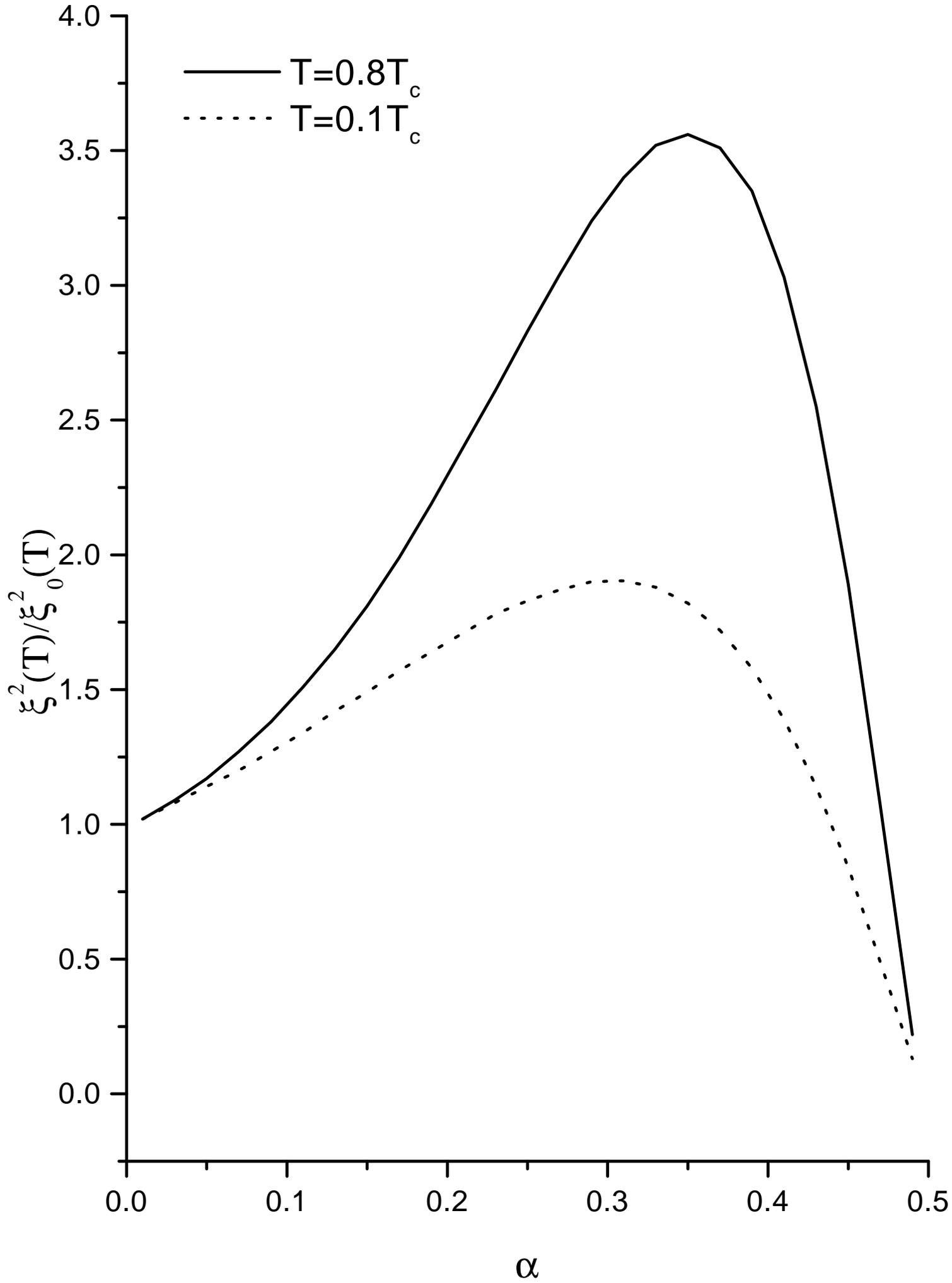
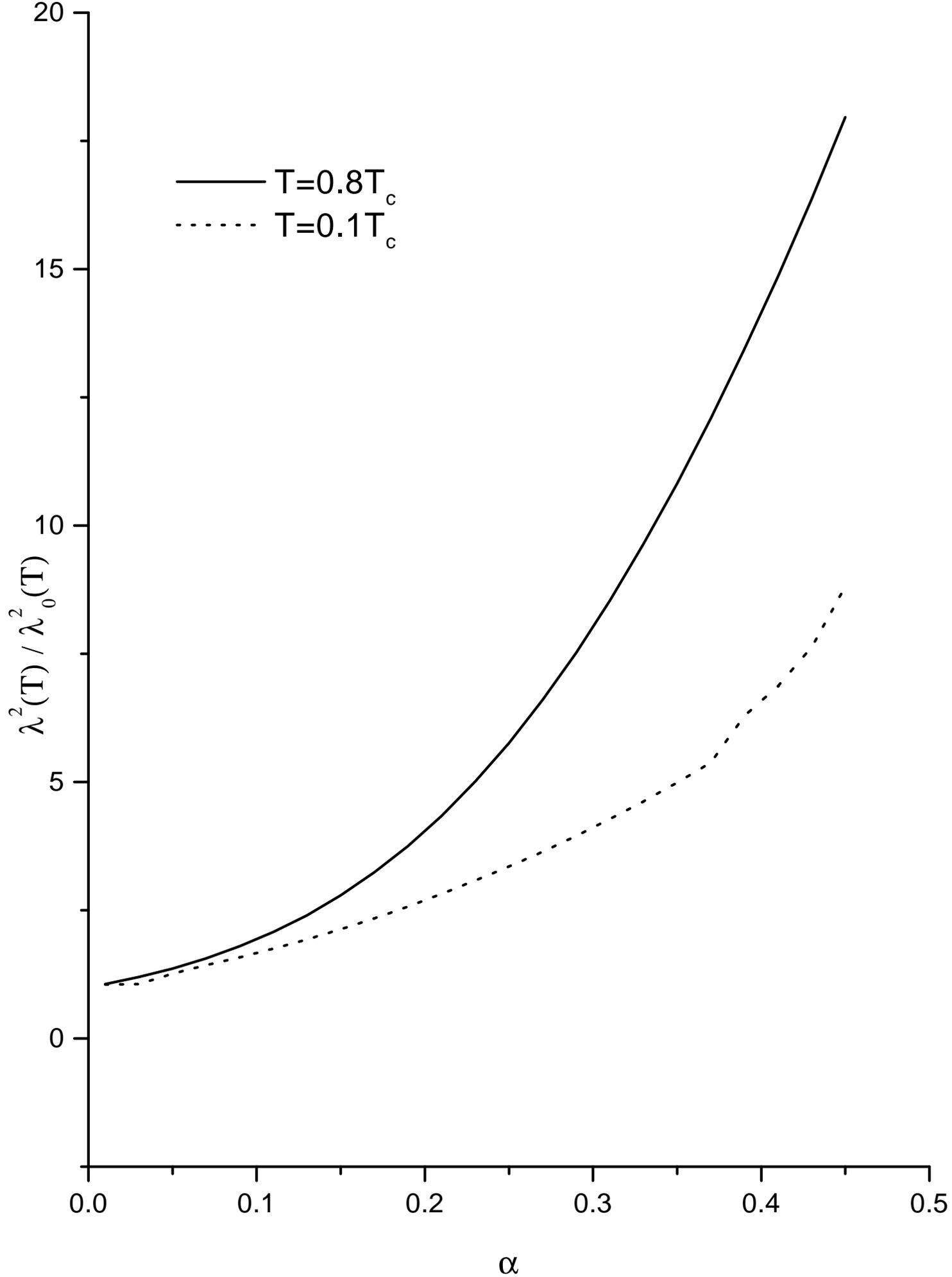
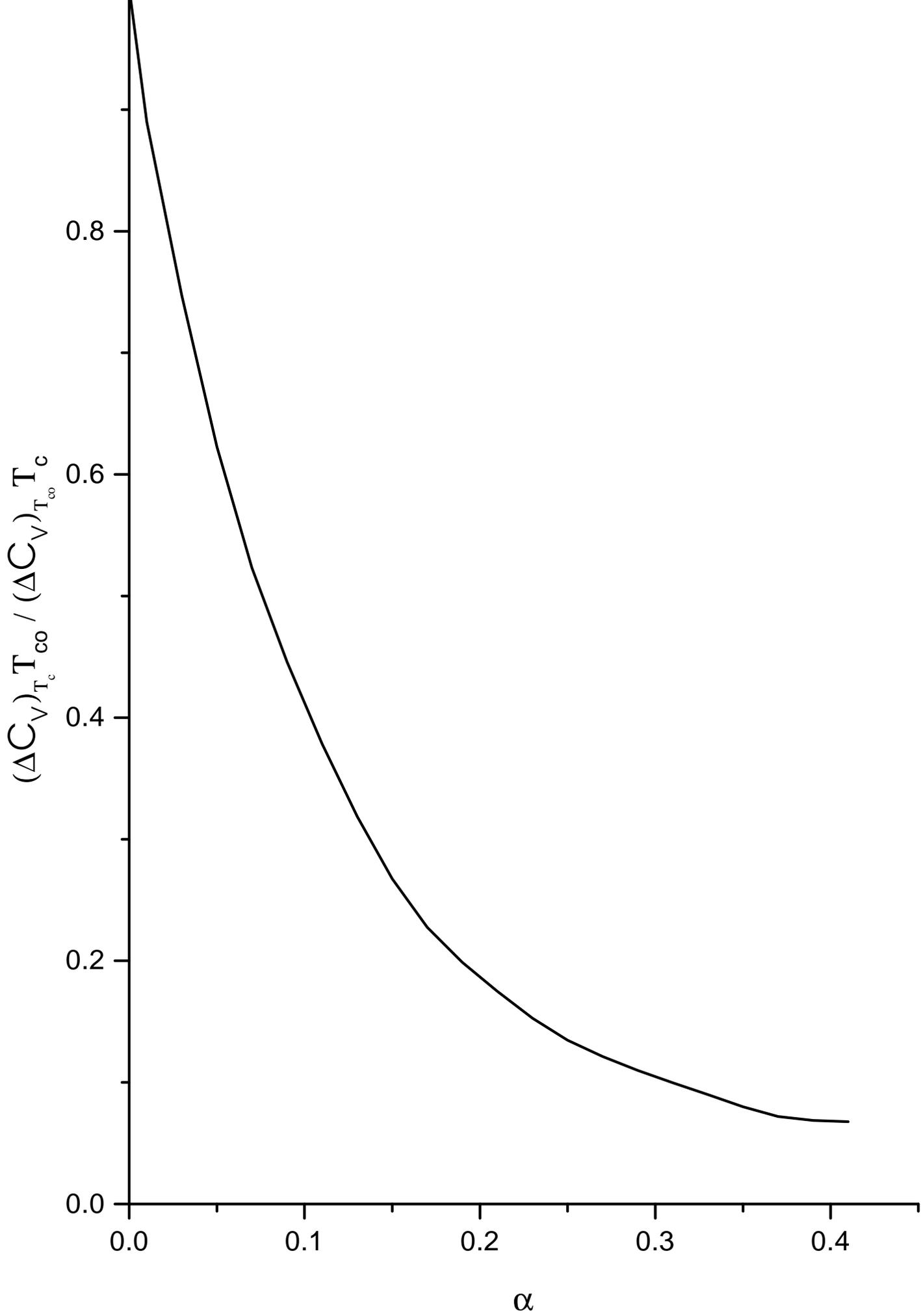
\end{document}